\documentclass{llncs}
\usepackage{times}
\usepackage[title]{appendix}
\usepackage{graphicx}
\usepackage{amsmath, amssymb}
\usepackage{multirow}
\usepackage{algpseudocode}
\usepackage{algorithmicx,algorithm}

\newcommand{\ATmodel}{\mathbf{AT}}

\newcommand{\STL}{\mathrm{STL}}
\newcommand{\SA}{\mathrm{SA}}
\newcommand{\CE}{\mathrm{CE}}
\newcommand{\AAAC}{\mathbf{A3C}}
\newcommand{\DQN}{\mathbf{DDQN}}

\DeclareMathOperator{\rob}{\rho}

\DeclareMathOperator{\hrz}{\mathsf{fr}}
\DeclareMathOperator{\reward}{\mathsf{reward}}
\DeclareMathOperator{\hst}{\mathsf{pr}}

\DeclareMathOperator*{\argmin}{arg\,min}
\DeclareMathOperator*{\argmax}{arg\,max}
\DeclareMathOperator{\UNTIL}{\mathcal{U}}
\DeclareMathOperator{\SINCE}{\mathcal{S}}

\algnewcommand\algorithmicinput{\textbf{input:}}
\algnewcommand\INPUT{\item[\algorithmicinput]}
\algnewcommand\algorithmicparameters{\textbf{parameters:}}
\algnewcommand\PARAMETERS{\item[\algorithmicparameters]}
\algnewcommand\algorithmicoutput{\textbf{output:}}
\algnewcommand\OUTPUT{\item[\algorithmicoutput]}

\title{Falsification of Cyber-Physical Systems Using Deep Reinforcement Learning}

\author{
Takumi Akazaki\inst{1}\inst{2} \and Shuang Liu\inst{3} \and Yoriyuki Yamagata\inst{4} \and Yihai Duan\inst{3} \and Jianye Hao\inst{3}
}

\institute{The University of Tokyo \\
\and
Japan Society for the Promotion of Science \\
\and
School of Software, Tianjin University \\
\and
National Institute of Advanced Industrial Science and Technology (AIST)
}

\begin{document}
\maketitle

\begin{abstract}
With the rapid development of software and distributed computing,  \emph{Cyber-Physical Systems} (CPS) are widely adopted in many application areas, e.g., smart grid, autonomous automobile. It is difficult to detect defects in CPS models due to the complexities involved in the software and physical systems. To find defects in CPS models efficiently, robustness guided falsification of CPS is introduced. Existing methods use several optimization techniques to generate counterexamples, which falsify the given properties of a CPS. However those methods may require a large number of simulation runs to find the counterexample and is far from practical. In this work, we explore state-of-the-art \emph{Deep Reinforcement Learning (DRL)} techniques
%, i.e., Asynchronous Advanced Acctor Critic (A3C) and Double Deep-Q Network (DDQN),
to reduce the number of simulation runs required to find such counterexamples. We %introduce our method and
report our method and the preliminary evaluation results.
\end{abstract}

%!TEX root = 0_main.tex
\section{Introduction}\label{sec:introduction}

\emph{Cyber-Physical Systems} (CPS) are more and more widely adopted in safety-critical domains, which makes it extremely important to guarantee the correctness of CPS systems.
% In model-based development, testing and verification are conducted on models before starting implementations, to detect defects early and avoid unnecessary costs.
Testing and verification on \emph{models} of CPS are common methods to guarantee the correctness.
However, it is hard for testing to achieve a high coverage; verification techniques are usually expensive and  undecidable~\cite{DBLP:conf/allerton/AbbasF12} due to the infinite state space
%(complexities involved in the software system, the physical systems as well as interactions between physical and software components)
of CPS models.
Therefore, robustness guided falsification~\cite{TLF_CPS_2013,DBLP:conf/tacas/AnnpureddyLFS11} method is introduced to detect defects efficiently.
In robustness guided falsification, \emph{Signal Temporal Logic} (STL)~\cite{bartocci2018specification} formulas are usually used to specify properties which must be satisfied by a CPS model.
Robustness of an STL formula, which is a numeric measure of how ``robust'' a property holds in the given CPS model, is defined.
The state space of the CPS model is explored and a trajectory which minimizes the robustness value is identified as a good candidate for testing.
%(which potentially leads to defects).
In this way, robustness guided falsification aids to generate defect-leading inputs (counterexamples), which enables more efficient, yet automatic detection of defects.
Although non-termination of robustness guided falsification does not mean the absence of counterexamples, it suggests the correctness of the CPS model to some extent.

Existing approaches adopt various kinds of stochastic global optimization algorithms e.g., simulated annealing~\cite{DBLP:conf/allerton/AbbasF12} and cross-entropy~\cite{DBLP:conf/hybrid/SankaranarayananF12}, to minimize robustness.
These methods take a full trajectory (a sequence of actions) as input, and adjusting input during the simulation is not supported. As a result, a large number of simulation runs are required in the falsification process.
Existing methods cannot guarantee finding a counterexample of practical CPS models in a limited time window because the simulation would then be tremendous.
% A smarter, knowledge guided method is required to accommodate faster feedback and accelerate the process of finding the counterexamples.

In this paper, we adopt \emph{deep reinforcement learning}   (DRL)~\cite{mnih2015human} algorithms to solve the problem of falsification of STL properties for CPS models.
Reinforcement learning techniques can observe feedbacks from the environment, and adjust the input action immediately. In this way, we are able to converge faster towards minimum robustness value.
%\ttodo{It becomse easy to read if we state
%  ``what is the environment, input action, and episode in our RL setting''.}
In particular, we adopt two state-of-the-art DRL techniques, i.e., \emph{Asynchronous Advanced Actor Critic} (A3C) and Double \emph{Deep-Q Network} (DDQN).
Our contributions are two folds:
(1) we show how to transform the problem of falsifying CPS models into a reinforcement learning problem; and
(2) we implement our method and conduct preliminary evaluations to show DRL technology can help reduce the number of simulation runs required to find a falsifying input for CPS models.
Reducing the number of simulation runs is important because during falsification, the majority of execution time is spent for simulation runs if CPS models are complex.

\subsubsection{Related Work}
There are two kinds of works, i.e., robustness guided falsification and controller synthesis, which are most related to our approach.

In robustness guided falsification methods, quantitative semantics
%called \emph{robust semantics}~\cite{DBLP:conf/fates/FainekosP06,DBLP:journals/tcs/FainekosP09}
over \emph{Metric Interval Temporal Logic} (MITL) and its variants STL~\cite{DBLP:conf/formats/MalerN04,DBLP:conf/formats/DonzeM10} are employed.  %Signal Temporal Logic (STL)
%Intuitively, these quantitative truth values stand for ``how robustly the formula is satisfied.''
Then the fault detection problem is translated into the numerical minimization problem.
%in 2 steps:
%(1) The parameterization scheme from the input space of the system is mapped into the finite dimensional Euclidian space $\mathbb{R}^m$. Typically, a continuous-time input signal $\sigma_\theta$ is represented by a finite sequence of its values $\theta \in \mathbb{R}^m$.
%(2) Pick a parameter $\theta \in \mathbb{R}^m$, simulate the system on the input signal $\sigma_\theta$, and measure the robustness of $\varphi$. Iterate the process and the next input parameter $\theta$ is chosen with the expectation of getting a smaller robustness value.
Several tools e.g., S-TaLiRo~\cite{DBLP:conf/tacas/AnnpureddyLFS11,DBLP:conf/cpsweek/HoxhaAF14a} and Breach~\cite{DBLP:conf/cav/Donze10} are developed to realize this approach.
%the robustness guided falsification approaches.
Moreover, various kind of numerical optimization techniques, e.g., simulated annealing~\cite{DBLP:conf/allerton/AbbasF12},
cross-entropy~\cite{DBLP:conf/hybrid/SankaranarayananF12}, and
Gaussian process optimization~\cite{DBLP:journals/corr/BartocciBNS13,DBLP:journals/tcs/BartocciBNS15,DBLP:conf/rv/Akazaki16,DBLP:conf/ifm/SilvettiPB17}, are studied
to solve the falsification problem efficiently.
% One benefit of robustness guided falsification approach is that the system is treated as a blackbox, which makes it applicable even if details of the system is complex and not well understood, which is common for CPS.
All these methods optimize the whole output trajectory of a CPS by changing the whole input trajectory.
As stated above, we use reinforcement learning which can observe feedbacks from a CPS and adjust the input immediately.
Thus, our method can be expected to arrive the falsifying input faster.

%% ========================================================================
%The other kind of approach is to apply the controller synthesis techniques.
In contrast to robustness guided falsification,
controller synthesis techniques enable choosing the input signal at a certain step based on observations of output signals.
There are works that synthesize the controller to enforce the \emph{Markov decision process}
to satisfy a given LTL formula~\cite{DBLP:conf/cdc/SadighKCSS14,DBLP:conf/wafr/LunaLMK14,DBLP:journals/tac/DingSBR14,DBLP:conf/hybrid/SoudjaniM17,DBLP:conf/cdc/DingSBR11}.
%% The important assumption for controller synthesis techniques is that
%% the target system satisfies Markov property.
The most closest related works~\cite{DBLP:conf/iros/LiVB17,DBLP:journals/corr/abs-1709-09611}
apply reinforcement learning techniques to enforce the small robotic system to satisfy the given LTL formula.
%TODO DRL and non-Markovian system are not related.  "Standard" theory of RL assumes the system Markovian and DRL is no different.  Rather, the difference of these works and our works are: our works falsify the properties while the control synthesis keep the properties (the difference is intent and application, rather than mathematical content.)  Another difference is we use DRL, which enables us to use complex non-linear functions for a Q-function or a policy function.
%The difference between these work and ours are as follows:
Our work is different from those works in two aspects: (1) we falsify the properties while the control synthesis methods try to satisfy the properties; and
(2) with DRL, we could employ complex non-linear functions to learn and model the environment, which is suitable to analyze the complex dynamics of CPS.
\section{Preliminary}\label{sec:preliminary}
%We briefly introduce the preliminary concepts used in our work.

%\vspace{4mm}
%\subsection{Metric Temporal Logic}
%\subsubsection{Robustness guided falsification}
\subsubsection{{Robustness guided falsification}}
In this paper, we employ a variant of \emph{Signal Temporal Logic ($\STL$)} defined in~\cite{bartocci2018specification}. The syntax is defined in the equation (\ref{eq:MTL}),
\begin{equation}\label{eq:MTL}
  \varphi ::= v \sim c \mid p \mid \neg \varphi \mid
  \varphi_1 \vee \varphi_2
  \mid \varphi_1 \UNTIL_I \varphi_2
  \mid \varphi_1 \SINCE_I \varphi_2
\end{equation}
where $v$ is \emph{real} variable,
$c$ is a rational number,
$p$ is atomic formula,
$\sim \in \{<, \leq\}$
and $I$ is an interval over non-negative real numbers.
%\ttodo{Propositional variables are omitted.
%  I think we need them to treat True in the definition of box operator,
%  and gears in our experiment.}
If $I$ is $[0, \infty]$, $I$ is omitted.
We also use other common abbreviations,
e.g., $\square_I \varphi \equiv \mathsf{True} \UNTIL_I \varphi$ and
$\boxminus_I \varphi \equiv \mathsf{True} \SINCE_I \varphi$.
For a given formula $\varphi$, an output signal $\mathbf{x}$ and time $t$, we adopt the notation of work~\cite{bartocci2018specification} and denote the \emph{robustness degree} of output signal $\mathbf{x}$ satisfying $\varphi$ at time $t$ by $\rob(\varphi, \mathbf{x}, t)$.
%Note that,
%intuitively,
%the robustness degree $\rob(\varphi, x, t)$
%stands for
%how ``robust'' the signal $x$ satisfies the formula $\varphi$ at time $t$.

%% Robustness~\cite{TLF_CPS_2013} of a MTL formula $\phi$ for a output trace $y$ at the time instant $t_n$ is defined as
%% \todo{Does n means $t_n$ in function rob()?}
%% $\rob(\mathbf{y}, n, \phi)$ , and is a measure of how
%% ``robust'' $\phi$ holds.
%% For atomic formula $p$, the robustness $\rob(\mathbf{y}, n, p)$ is defined as the infimum of the distance of the point $y$ which does not satisfies $p$ from $y_n$.
%% \todo{$y_n$ means $y_t$? or $y_tn$? We either use t or n, be consistent}
%% %The distance between two states $\dist(y, y_n)$ can be any metric, but
%% \textcolor{red}{
%% In this paper we use Euclidian metric to define the distance between two states $\dist(y, y_n)$.
%% For example, if $y_n = 0$ and $p = \{ y \mid y < 1 \}$, $\rob(\mathbf{y}, n, p) = \dist(y_n, y) = 1$.}
%% %

We also adopt the notion of \emph{future-reach} $\hrz(\varphi)$ and
\emph{past-reach} $\hst(\varphi)$ following~\cite{DBLP:conf/rv/HoOW14}.
%% The \emph{horizon} $\hrz(\phi)$ of a MTL formula $\phi$ is the time in future which is required to determine the truth value of the formula $\phi$.
%Generally speaking, for a given formula $\varphi$,
Intuitively, $\hrz(\varphi)$ is the time in future which is required to determine the truth value of formula $\varphi$, and $\hst(\varphi)$ is the time in past.
For example,
$\hrz(p) = 0$, $\hrz(\square_{[0, 3]}p) = 3$ and $\hrz(\boxminus_{[0,3]}p) = 0$.
%%Similarly we define the \emph{history} $\hst(\varphi)$.
Similarly, for past-reach,
$\hst(p) = 0$, $\hst(\square_{[0, 3]}p) = 0$, $\hst(\boxminus_{[0,3]}p) = 3$.

%% \begin{lemma}[Robustness of a past dependent formula]\label{lem:rob-past}
%%   Let $\mathbf{y}$ be a finite trace of system states $y_0, \ldots, y_n$.
%%   Let $\overline{\mathbf{y}}_1$ and $\overline{\mathbf{y}}_2$ be two infinite extensions of $\mathbf{y}$.
%%   If $\phi$ is past dependent,
%%   \begin{equation}
%%     \rob(\overline{\mathbf{y}}_1, t, \phi) = \rob(\overline{\mathbf{y}}_2, t, \phi)
%%   \end{equation}
%% \end{lemma}

%% By Lemma \ref{lem:rob-past}, robustness of a past dependent formula at instant $n$ is completely determined by $y_0, \ldots, y_n$.
%% Therefore, we use the notation $\rob(\mathbf{y}, t, \phi)$ for robustness of a past-dependent formula $\phi$ on a finite trace $\mathbf{y}$.

%\subsection{Past-dependent life-long property falsification}
In this paper, we focus on a specific class of the temporal logic formula called \emph{life-long property}.
%to employ our approach.

\begin{definition}[life-long property]
  A \emph{life-long property} is an $\STL$ formula $\psi \equiv \square \varphi$ where $\hrz(\varphi),
  \hst(\varphi)$ are finite.
  If $\hrz(\varphi) = 0$, we call $\psi$ \emph{past-dependent life-long property}.

\end{definition}

%% %TODO life-long property, not past-dependent
%% \begin{definition}[Past-dependent life-long property]
%%   A \emph{life-long property} is an $\mathsf{MTL}$ formula $\psi \equiv \square \varphi$ where $\varphi$ only has a finite horizon.
%%   In particular, if $\varphi$ only contains bounded modal operators, $\psi$ is a life-long property.
%%   If $\varphi$ is past-dependent, then $\square \varphi$ is called \emph{past-dependent life-long property}.
%% \end{definition}

%\vspace{4mm}
\subsubsection{Reinforcement Learning}
%\subsection{Reinforcement learning}
%\subsubsection{Reinforcement Learning}
Reinforcement learning is one of machine learning techniques in which an agent learns the structure of the environment based on observations, and maximizes the rewards by acting according to the learnt knowledge.
% Reinforcement learning is first proposed and used in the domain of audio and image processing to improve the analysis performance. Reinforcement learning has shown its power and potential in training AlphaGo Zero~\cite{AlphaGo0}, which became the world's best Go player in 40 days, from scratch.
%In this work, we are using reinforcement learning techniques to reduce the accelerate the process of finding the counterexample, which falsifies the robustness property defined for a CPS.
%In particular, we adopt Asynchronous Advantage Actor-Critic (A3C) and Double Deep Q Network (DDQN) in our problem.
%
%Fig. \ref{fig:RL} shows the standard setting of reinforcement learning.
%%
%
%\begin{figure}
%  \centering
%  \scriptsize
%  \includegraphics[scale=0.67]{fig/RL.pdf}
%  \caption{Reinforcement learning setting}
%  \label{fig:RL}
%  \vspace{-7mm}
%\end{figure}
%
%%
The standard setting of a reinforcement learning problem consists of an agent and an environment. %, as shown in Fig.~\ref{fig:arch}.
The agent observes the current state and reward from the environment, and returns the next action to the environment.
%
%%
%%%
% Reinforcement learning is often formulated as a \emph{Markov decision process (MDP)}~\cite{Szepesvari2010}.
% A MDP is a triple $\mathcal M = (\mathcal X, \mathcal A, \mathcal P_0)$.
% \state is a set of states, \action is a set of actions and \probkernal is the transition probability kernel.
% A transition probability kernel \probkernal assigns a probability distribution (over $\mathcal X \times \mathbb R$, which is a distribution over the next states and the reward when the agent takes an action $a$ at the state $x$.), to each state-action pair $(x, a) \in \mathcal X \times \mathcal A$.
The goal of reinforcement learning is for each step $n$, given the sequence of previous states $x_0, \ldots, x_{n-1}$, rewards $r_1, \ldots, r_{n}$ and actions $a_0, \ldots, a_{n-1}$, generate an action $a_n$, which maximizes expected value of the sum of rewards:
%\begin{equation}
 $ r = \sum_{k = n}^\infty \gamma^k r_{k+1}$
%\end{equation}
, where $0 < \gamma \leq 1$ is a discount factor.
Deep reinforcement learning is a reinforcement learning technique which uses a \emph{deep neural network} for learning.  % to represent a $Q$-function and/or a policy $\pi$.
In this work, we particularly adopted 2 state-of-the-art deep reinforcement learning algorithms, i.e., \emph{Asynchronous Advantage Actor-Critic} (A3C)~\cite{Mnih2016} and \emph{Double Deep Q Network} (DDQN)~\cite{pmlr-v48-gu16}.
\section{Our Approach}\label{sec:overview}

\begin{algorithm}[tp]
\scriptsize
  \caption{Falsification for $\psi = \square \varphi$ by reinforcement learning}
  \label{algo:RLfalsification}
  \begin{algorithmic}[1]
    \INPUT A past-dependent life-long property $\psi = \square \varphi$, a system $\mathcal{M}$,
    an agent $\mathcal{A}$
    \OUTPUT A counterexample input signal $\mathbf{u}$ if exists
    \PARAMETERS A step time $\Delta_T$, the end time $T_{\mathsf{end}}$, the maximum number of the episode $N$
 %   \Ensure something
    \For{$\mathsf{numEpisode} \gets$ $1$ to $N$}
    \State $i \gets 0$, $r \gets 0$, $x$ be the initial (output) state of $\mathcal{M}$
    \State $\mathbf{u}$ be the empty input signal sequence
    \While{$i \Delta_T < T_{\mathsf{end}}$}
    \State $u \gets \mathcal{A}.\mathsf{step}(x, r)$,  $\mathbf{u} \gets \mathsf{append}(\mathbf{u}, (i \Delta_T, u))$
    \Comment choose the next input by the agent
    \State $\mathbf{x} \gets \mathcal{M}(\mathbf{u})$, $x \gets \mathbf{x}((i+1)\Delta_T)$
    \Comment simulate, observe the new output state
    \State $r \gets \reward(\mathbf{x}, \psi)$
    \State $i \gets i+1$
    \Comment calculate the reward by following eq.~(\ref{def:reward})
    \EndWhile
    \If{$\mathbf{x} \not\models \psi$}
    %% \Then
    \Return $\mathbf{u}$ as a falsifying input
    \EndIf
    \State $\mathcal{A}.\mathsf{reset}(x, r)$
    \EndFor
   \end{algorithmic}
\end{algorithm}
%
% In this section, we describe our method
% of enforcing the CPS model to falsify the given STL specification
% by reinforcement learning.

\subsection{Overview of our algorithm}\label{subsec:algorithm}
Let us consider the falsification problem to find a counterexample of the life-long property $\psi \equiv \square \varphi$.
If the output signal is infinitely long to past and future directions, $\psi$ is logically equivalent to a past-dependent life-long property $\square \boxminus_{[\hrz(\varphi), \hrz(\varphi)]} \varphi$.
In general, the output signal is not infinitely long to some direction but using this conversion we convert all life-long properties to past-dependent life-long properties.
Our evaluation in Section \ref{sec:exp} suggests that this approximation does not adversely affect the performance.
Therefore, assume $\psi$ is a past-dependent life-long property, we generate an input signal $\mathbf{u}$ for system $\mathcal{M}$,
such that the corresponding output signal $\mathcal{M}(\mathbf{u})$ does not satisfy $\psi$.

In our algorithm,
we fix the simulation time to be $T_{\mathsf{end}}$
and
call one simulation until time $T_\mathsf{end}$
an \emph{episode} in conformance with the reinforcement learning terminology.
We fix the discretization of time to a positive real number $\Delta_T$.
%For an agent $\mathcal{A}$,
%in each episode,
%it generates an input signal $\mathbf{u}(t)$
%%adaptively
%based on the observed current system output and the reward.
%More precisely,
The agent $\mathcal{A}$ generates the piecewise-constant input signal
$\mathbf{u} = \big[(0, u_0), (\Delta_T, u_{1}), (2\Delta_T, u_{2}), \dots \big]$
by iterating the following steps:

%\begin{enumerate}
%\item
% At time $i \Delta_T$ ($i=0,1,\dots$),
%  the agent $\mathcal{A}$ choose the next input value $u_i$.
%  The generated input signal is extended to
%  $\mathbf{u} = \big[(0, u_0), \dots, (i\Delta_T, u_i) \big]$. \\
%\item
%  Our algorithm obtains the corresponding output signal $\mathbf{x} = \mathcal{M}(\mathbf{u})$
%  by stepping forward one simulation on the model $\mathcal{M}$
%  from time $i \Delta_T$ to $(i+1) \Delta_T$ with input $u_{i}$. \\
%\item
% Let $x_{i+1} = \mathbf{x}((i+1)\Delta_T)$ be the new (observed) state (i.e., output) of the system. \\
%\item
% We compute reward $r_{i+1}$ by $\reward(\varphi, \mathbf{x}, (i+1)\Delta_T)$ (defined in Section \ref{subsec:reward}). \\
%\item
% $\mathcal{A}$ updates its action based on the new state $x_{i+1}$ and the reward $r_{i+1}$.
%\end{enumerate}

(1) At time $i \Delta_T$ ($i=0,1,\dots$),
  the agent $\mathcal{A}$ chooses the next input value $u_i$.
  The generated input signal is extended to
  $\mathbf{u} = \big[(0, u_0), \dots, (i\Delta_T, u_i) \big]$. \\
%\item
\indent (2) Our algorithm obtains the corresponding output signal $\mathbf{x} = \mathcal{M}(\mathbf{u})$
  by stepping forward one simulation on the model $\mathcal{M}$
  from time $i \Delta_T$ to $(i+1) \Delta_T$ with input $u_{i}$. \\
%\item
\indent (3) Let $x_{i+1} = \mathbf{x}((i+1)\Delta_T)$ be the new (observed) state (i.e., output) of the system. \\
%\item
\indent (4) We compute reward $r_{i+1}$ by $\reward(\varphi, \mathbf{x}, (i+1)\Delta_T)$ (defined in Section \ref{subsec:reward}). \\
%\item
\indent (5) The agent $\mathcal{A}$ updates its action based on the new state $x_{i+1}$ and reward $r_{i+1}$.

At the end of each episode,
we obtain the output signal trajectory $\mathbf{x}$,
and check whether it satisfies the property $\psi = \square \varphi$ or not.
If it is falsified, return the current input signal $\mathbf{u}$ as a counterexample.
Otherwise, we discard the current generated signal input
and restart the episode from the beginning.

The complete algorithm of our approach is shown in Algorithm~\ref{algo:RLfalsification}.
The method call $\mathcal{A}.\mathsf{step}(x, r)$ represents
the agent $\mathcal{A}$ push the current state reward pair ($x$, $r$) into its memory
and returns the next action $u$ (the input signal in the next step).
The method call $\mathcal{A}.\mathsf{reset}(x, r)$ notifies the agent that the current episode is completed, and returns the current state and reward.
%\ttodo{This sentence seems misleadning as is discussed in our skype chat.}
Function $\reward(\mathbf{x}, \psi)$ calculates the reward based on Definition~\ref{def:reward}.

\subsection{Reward definition for life-long property falsification}\label{subsec:reward}
Our goal is to find the input signal $\mathbf{u}$ to the
system $\mathcal{M}$ which minimizes $\rob(\psi, \mathcal{M}(\mathbf{u}), 0)$ where $\psi = \square \varphi$ and $\rho$ is a robustness.
We determine $u_0, u_1, \ldots$ in a greedy way.
Assume that $u_0, \ldots, u_i$ are determined.
$u_{i+1}$ can be determined by
\begin{align}
\scriptstyle
\label{eq:action}
  u_{i+1} &= \argmin_{u_{i+1}} \min_{u_{i+2}, \ldots} \rob(\square \varphi, \mathcal{M}(\left[(0, u_0), (\Delta_T, u_1), \ldots \right]), 0) \\
  &\sim \argmax_{u_{i+1}} \max_{u_{i+2}, \ldots} \sum_{k=i+1}^\infty \{ e^{- \rob(\varphi, \mathcal{M}(\left[(0, u_0), \ldots, (k\Delta_T, u_k) \right]), k\Delta_T)} - 1\} \label{eq:r}
\end{align}
The detailed derivation steps can be found in Appendix~\ref{sec:appendix}.
%\eqref{eq:disc} uses the fact $\varphi$ is past-dependent and \eqref{eq:logsum} uses an approximation of minimum by the log-sum-exp function~\cite{cook2011basic}.

In our reinforcement learning base approach, we use discounting factor $\gamma=1$ and reward $r_i = e^{- \rob(\varphi, \mathcal{M}(\left[(0, u_0), \ldots, (i\Delta_T, u_i)\right]), i\Delta_T)} - 1$  to approximately compute action $u_{i+1}$, from $u_0, \ldots, u_i$, $\mathcal{M}(\left[(0, u_0), \ldots, (i\Delta_T, u_i) \right])$ and $r_1, \ldots, r_i$.
%\ttodo{No guarantee that u is approximately computed.
%  We cannot estimate the approximation error.
%  I prefer we claim
%  ``we use the discounting factor and reward to hopefully compute the approximation of the next action u...''
%}
%If the reward $r_i = e^{- \rob(\varphi, \mathcal{M}(\left[(0, u_0), \ldots, (i\Delta_T, u_i)\right]), i\Delta_T)} - 1$ and discounting factor $\gamma=1$ are used, we expect a reinforcement learning algorithm
%approximately computes $u_{i+1}$ as an action from $u_0, \ldots, u_i$, $\mathcal{M}(\left[(0, u_0), \ldots, (i\Delta_T, u_i) \right])$ and $r_1, \ldots, r_i$.

\begin{definition}[reward]\label{def:reward}
  Let $\psi \equiv \square \varphi$ be a past-dependent formula
  and $\mathbf{x} = \mathcal{M}(\mathbf{u})$ be a finite length signal until the time $t$.
  We define the reward $\reward(\psi, \mathbf{x})$ as
  \begin{equation}\label{eq:reward}
    \reward(\psi, \mathbf{x}) =
      \exp(- \rob(\varphi, \mathbf{x}, t)) - 1
  \end{equation}
\end{definition}

%!TEX root = 0_main.tex
\section{Preliminary Results}\label{sec:exp}
%To evaluate the efficiency and effectiveness of our work, we conduct experiments with well known CPS models in Matlab/Simulink.
%We compare our reinforcement learning based technique with existing methods, i.e., simulated annealing and cross entropy based methods, and analyze the results.
%\ytodo{Mention Breach when we finish the comparison with Breach}

%We discuss our implementation and report our preliminary evaluation result in this paper.

%\subsection{Implementation}
%\vspace{5mm}
\noindent\textbf{Implementation}
%\subsubsection{Implementation}
\begin{figure}[tp]
\centering
\scriptsize
\includegraphics[scale=0.44]{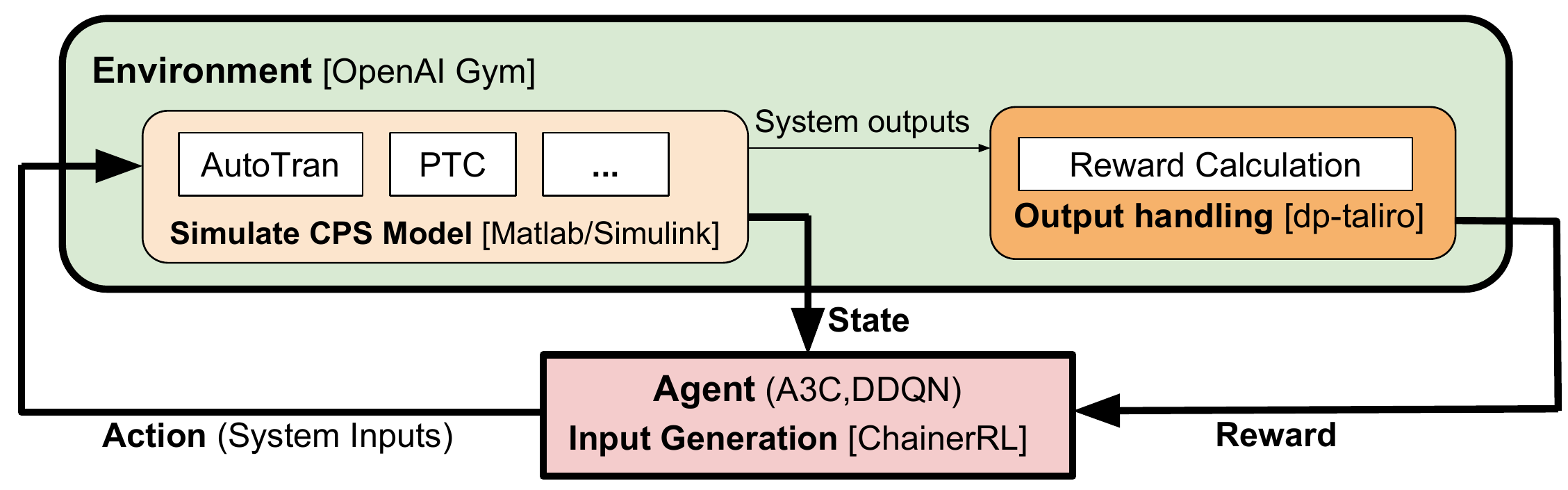}
\caption{Architecture of our system}
\label{fig:arch}
\vspace{-7mm}
\end{figure}
The overall architecture of our system is shown in Fig.~\ref{fig:arch}.
Our implementation consists of three components, i.e., input generation, output handling and simulation.
The input generation component adopts reinforcement learning techniques and is implemented based on the ChainerRL library~\cite{ChainerRL}.
We use default hyper-parameters in the library or sample programs without change.
The output handling component conducts reward calculation using dp-TaliRo~\cite{S-TaliRo}.
The simulation is conducted with Matlab/Simulink models, which are encapsulated by the openAI gym library~\cite{1606.01540}.
\vspace{4mm}
\noindent \textbf{Evaluation Settings}
%\subsubsection{Evaluation Settings}
We use a widely adopted CPS model, automatic transmission control system ($\ATmodel$) ~\cite{bardh2014benchmarks}, to evaluate our method.
%
%The system model is from a public demonstration of modeling an automatic transmission control with the Stateflow~\cite{ATinSF}
%%~\footnote{https://mathworks.com/products/stateflow.html}
%package in MATLAB.
%
$\ATmodel$ has throttle and brake as input ports, and the output ports are the vehicle velocity $v$, the engine rotation speed $\omega$ and the current gear state $g$.
%Although the size of the model is relatively small comparing the actual systems in industry, but its dynamics contains both discrete and continuous values.
%Therefore, it is suitable as a benchmark of falsification on CPSs.
%
We conduct our evaluation with the formulas in Table~\ref{tab:formulas}.
Formulas $\varphi_1$--$\varphi_6$ are rewriting of  $\varphi^{AT}_1$--$\varphi^{AT}_6$ in benchmark~\cite{bardh2014benchmarks} into life-long properties in our approach.
%We do not use the original benchmark because we focus on life-long properties.
In addition, we propose three new formulas $\varphi_{7}$--$\varphi_{9}$.
%\textcolor{red}{For all formulas, we tune parameters in the formulas such that they are difficult to falsify, therefore can differentiate performance of each method.}
%
For each formula $\varphi_1$--$\varphi_9$, we compare the performance of our approaches (A3C, DDQN), with the baseline algorithms, i.e., simulated annealing ($\SA$) and cross entropy ($\CE$).
For each property, we run the falsification procedure 20 times.
For each falsification procedure, we execute simulation episodes up to 200 times and measure the number of simulation episodes required to falsify the property.
If the property cannot be falsified within 200 episodes, the procedure fails.
%We record whether each falsification procedure is successful or not.
%
% Further, for fair comparison, we change sampling rate and choose the best performing rate for each combination of a method and formula.
% Here, the best means the smallest median of the number of simulations runs required to falsify the formula.
% If the tie occurs, we further compares success rates of the falsification procedure.
% Finally, if still we cannot decide we use the media of the execution time of the falsification processes.
%
We observe that $\Delta_{T}$ may strongly affect the performance of each algorithm.
%For example, simulated annealing tends to perform badly if we use high sampling rate.
%On the other hand, reinforcement learning based methods are not affected by high sampling rate.
%Based on the above observation,
Therefore, we
%choose the $\Delta_{T}$ which gives the best performance (prioritized by numEpisode and success rate) of each algorithm.
vary $\Delta_{T}$ (among \{1, 5, 10\} except for the cases of $\AAAC$ and $\DQN$ for $\varphi_7$--$\varphi_9$ among we use \{5, 10\}~\footnote{These methods with $\Delta_{T}=1$ for $\varphi_7$--$\varphi_9$ shows bad performance and did not terminate in 5 days.}) and report the setting (of $\Delta_{T}$) which leads to the best performance (the least episode number and highest success rate) for each algorithm.
%We change $\Delta_{T}$ from 1, 5, 10 for $\varphi_1$--$\varphi_6$ and 5, 10 for $\varphi_7$--$\varphi_9$.
%$\Delta_{T}$ 1 for $\varphi_5$--$\varphi_7$ is omitted because of the performance reason.

\begin{table}[tp]
  \centering
  \begin{minipage}[t]{.48\textwidth}
    \centering
    \scriptsize
    \begin{tabular}{c||c}
      id & Formula\\
      \hline
      \hline
      $\varphi_1$ & $\square \omega \leq \overline{\omega}$\\
      $\varphi_2$ & $\square (v \leq \overline{v} \wedge \omega \leq \overline{\omega})$\\
      $\varphi_3$
      & $\square ((g_2 \wedge \diamond_{[0, 0.1]} g_1) \rightarrow \square_{[0.1, 1.0]} \neg g_2)$\\
      $\varphi_4$
      & $\square ((\neg g_1 \wedge \diamond_{[0, 0.1]} g_1) \rightarrow \square_{[0.1, 1.0]} g_1)$\\
      $\varphi_5$
      & $\square \bigwedge_{i=1}^4 ((\neg g_i \wedge \diamond_{[0, 0.1] g_i}) \rightarrow \square_{[0.1, 1.0]} g_i)$\\
    \end{tabular}
  \end{minipage}
  \begin{minipage}[t]{.48\textwidth}
    \centering
    \scriptsize
    \begin{tabular}{c||c}
      id & Formula\\
      \hline
      \hline
      $\varphi_6$
      & $\square (\square_{[0, t_1]} \omega \leq \overline\omega \rightarrow \square_{[t_1, t_2]} v \leq \overline{v})$\\
      $\varphi_7$
      & $\square v \leq \overline{v}$\\
      $\varphi_8$
      & $\square \diamond_{[0,25]} \neg (\underline{v} \leq v \leq  \overline{v})$\\
      $\varphi_9$
      & $\square \neg \square_{[0,20]} (\neg g_4 \wedge \omega \geq \overline{\omega})$\\
    \end{tabular}
  \end{minipage}
  \caption{The list of the evaluated properties on $\ATmodel$.}
  \label{tab:formulas}
  \vspace{-5mm}
\end{table}
\begin{table}[tp]
  \centering
    \centering
    \scriptsize
    \begin{tabular}[t]{c||c c c c|c c c c|c c c c|}
      id & \multicolumn{4}{|c|}{$\Delta_T$} & \multicolumn{4}{|c|}{Success rate} & \multicolumn{4}{|c|}{$\mathsf{numEpisode}$}\\
      \hline
      & $\AAAC$ & $\DQN$ & $\SA$ & $\CE$ & $\AAAC$ & $\DQN$ & $\SA$ & $\CE$ & $\AAAC$ & $\DQN$ & $\SA$ & $\CE$ \\
      \hline
      \hline
      $\varphi_1$ & 5 & 1 & 10 & 5 & $\textbf{100}\%^*$ & $\textbf{100}\%^*$ & 65.0\% & 10.0\% & $\textbf{16.5}^{**}$ & 24.5 & 118.5 & 200.0\\
      $\varphi_2$ & 5 & 1 & 10 & 5 & $\textbf{100}\%^*$ & $\textbf{100}\%^*$ & 65.0\% & 10.0\% & $\textbf{11.5}^{**}$ & 27.5 & 118.5 & 200.0\\
      $\varphi_3$ & 1 & 1 & 1 & 1 & 75.0 & 5.0\% & 20.0\% & \textbf{85.0}\% & 44.0 & 200.0 & 200.0 & \textbf{26.5}\\
      $\varphi_4$ & 1 & 1 & 1 & 1 & 75.0 & 10.0\% & 20.0\% & \textbf{85.0}\% & 67.5 & 200.0 & 200.0 & $\textbf{26.5}^{*}$\\
      $\varphi_5$ & 1 & 1 & 1 & 1 & \textbf{100}\% & \textbf{100}\% & \textbf{100}\% & \textbf{100}\% & \textbf{1.0} & 2.0 & \textbf{1.0} & \textbf{1.0} \\
      $\varphi_6$ & 10 & 10 & 10 & 10 & $\textbf{100}\%^*$ & $\textbf{100}\%^*$ & 70.0\% & 50.0\% & $\textbf{3.5}^{**}$ & $\textbf{3.5}^{**}$ & 160.5 & 119.0\\
      $\varphi_7$ & 5 & 5 & 1 & 1 & 65.0\% & $\textbf{100}\%^{**}$ & 0.0\% & 0.0\% & 125.0 & $\textbf{63.0}^{**}$ & 200.0 & 200.0\\
      $\varphi_8$ & 10 & 10 & 10 & 1 & 80.0\% & \textbf{95.0}\% & 90.0\% & 75.0\% & 72.0 & 52.0 & 83.0 & \textbf{21.0}\\
      $\varphi_9$ & 10 & 10 & 10 & 10 & 95.0\% & $\textbf{100}\%^{**}$ & 15.0\% & 5.0\% & 46.0 & $\textbf{12.0}^{**}$ & 200.0 & 200.0 \\
      \hline
  \end{tabular}
  \caption{The experimental result on $\ATmodel$.}
  \label{tab:ARCH2014}
   \vspace{-5mm}
\end{table}

%\subsection{Evaluation Results}\label{sec:result}
\vspace{4mm}
\noindent \textbf{Evaluation Results}
%\subsubsection{Evaluation Results}
The preliminary results are presented in Table.~\ref{tab:ARCH2014}.
%The algorithms indicated by bold face are our approach and others are baselines.
%%
%
%$\SA$ is simulated annealing and $\CE$ is cross entropy method.
%
%%
The $\Delta_{T}$ columns indicate the best performing $\Delta_{T}$ for each algorithm.
The ``Success rate'' columns indicate the success rate of falsification process.
The ``numEpisode'' columns show the median (among the 20 procedures) of the number of simulation episodes required to falsify the formula.
If the falsification procedure fails, we consider the number of simulation episodes to be the maximum allowed episodes (200).
We use median since the distribution of the number of simulation episodes tends to be skewed.
%Another reason is that our sample size is relatively small so that we need to avoid the effects of outliers.

The best results (success rate and numEpisode) of each formula are highlighted in bold.
If the difference between the best entry of our methods and the best entry of the baseline methods is statistically significant by Fisher's exact test and the Mann Whitney U-test~\cite{corder2014nonparametric}, we mark the best entry with $*$ ($p < 0.05$) or $**$ ($p < 0.001$), respectively.
%\todo{This is again a little misleading.  We do not compare, for example, A3C and DDQN.  Only between the best one among our method and the best one among baselines.}
%To test the success rate, we use Fisher's exact test of independence~\cite{corder2014nonparametric}.
%To test iteration numbers, we use the Mann-Whitney U-test~\cite{corder2014nonparametric}.

As shown in Table~\ref{tab:ARCH2014}, RL based methods almost always outperforms baseline methods on success rate, which means RL based methods are more likely to find the falsified inputs with a limited number of episodes.
This is because RL based methods learn knowledge from the environment and generate input signals adaptively during the simulations.
%On the other hand, the result on iteration numbers is more mixed.
Among the statistically significant results of numEpisode, our methods are best for five cases ($\varphi_1, \varphi_2,\varphi_6,\varphi_7,\varphi_9$), while the baseline methods are best for one case ($\varphi_4$).
%Although, for $\varphi_3$--$\varphi_5$ and $\varphi_8$ the cross entropy method is best in episode numbers, only in one case the result is statistically significant.
For the case of $\varphi_4$, it is likely because that
all variables in this formula take discrete values,
thus, reinforcement learning is less effective.
%the reward in reinforcement learning tends to be constant.
%\todo{Still A3C performs not so bad.  Also what is the reason of good performance of CE?}
Further, DDQN tends to return extreme values as actions,
which are not solutions to falsify $\varphi_3$ and $\varphi_4$.
This explains poor performance of DDQN for the case of $\varphi_3$ and $\varphi_4$.

\section{Conclusion and Future Work}\label{sec:conclusion}
In this paper, we report an approach which adopts reinforcement learning algorithms to solve the problem of robustness-guided falsification of CPS systems. We implement our approach in a prototype tool and conduct preliminary evaluations with a widely adopted CPS system. The evaluation results show that our method can reduce the number of episodes to find the falsifying input. As a future work, we plan to extend the current work to explore more reinforcement learning algorithms and evaluate our methods on more CPS benchmarks. 
\bibliographystyle{abbrv}
\bibliography{ref}

\begin{appendix}
%!TEX root = 0_main.tex
\setcounter{equation}{0}
\renewcommand{\theequation}{A\arabic{equation}}
\section{Derivation of Equation~(\ref{eq:r}) in Section \ref{sec:overview}}
\label{sec:appendix}
In Section \ref{sec:overview}, the derivation of \eqref{eq:r} from \eqref{eq:action} is omitted.
In this Appendix, the omitted derivation is presented.
\begin{align}
  &\phantom{=} \argmin_{u_{i+1}} \min_{u_{i+2}, \ldots} \rob(\square \varphi, \mathcal{M}(\left[(0, u_0), (\Delta_T, u_1), \ldots \right]), 0) \tag{\ref{eq:action}} \\
  &= \argmin_{u_{i+1}} \min_{u_{i+2}, \ldots} \min_{t \in \mathbb R} \rob(\varphi, \mathcal{M}(\left[(0, u_0), (\Delta_T, u_1), \ldots \right]), t)\\
  &\sim \argmin_{u_{i+1}} \min_{u_{i+2}, \ldots} \min_{k = i+1, i+2, \ldots} \rob(\varphi, \mathcal{M}(\left[(0, u_0), \ldots, (k\Delta_T, u_k) \right]), k\Delta_T) \label{eq:disc}\\
  &\sim \argmin_{u_{i+1}} \min_{u_{i+2}, \ldots} \left[ - \log \left\{ 1 + \sum_{k=i+1}^\infty \{ e^{- \rob(\varphi, \mathcal{M}(\left[(0, u_0), \ldots, (k\Delta_T, u_k) \right]), k\Delta_T)} - 1\} \right\} \right] \label{eq:logsum}\\
  &= \argmax_{u_{i+1}} \max_{u_{i+2}, \ldots} \sum_{k=i+1}^\infty \{ e^{- \rob(\varphi, \mathcal{M}(\left[(0, u_0), \ldots, (k\Delta_T, u_k) \right]), k\Delta_T)} - 1\} \tag{\ref{eq:r}}
\end{align}
\eqref{eq:action} is the equation which derives the next input $u_{i+1}$.
\eqref{eq:disc} uses the fact $\varphi$ is past-dependent and \eqref{eq:logsum} uses an approximation of minimum by the log-sum-exp function~\cite{cook2011basic}.
Finally, the last equation \eqref{eq:r} is the same equation \eqref{eq:r} in Section \ref{sec:overview}.
This completes the derivation.

\end{appendix}
\end{document}